\documentstyle[psfig]{mn}

\newif\ifAMStwofonts

\ifoldfss
  \ifCUPmtlplainloaded \else
    \NewTextAlphabet{textbfit} {cmbxti10} {}
    \NewTextAlphabet{textbfss} {cmssbx10} {}
    \NewMathAlphabet{mathbfit} {cmbxti10} {} 
    \NewMathAlphabet{mathbfss} {cmssbx10} {} 
  \fi
  \ifAMStwofonts
    \ifCUPmtlplainloaded \else
      \NewSymbolFont{upmath} {eurm10}
      \NewSymbolFont{AMSa} {msam10}
      \NewMathSymbol{\upi}     {0}{upmath}{19}
      \NewMathSymbol{\umu}     {0}{upmath}{16}
      \NewMathSymbol{\upartial}{0}{upmath}{40}
      \NewMathSymbol{\leqslant}{3}{AMSa}{36}
      \NewMathSymbol{\geqslant}{3}{AMSa}{3E}

      \let\leq=\leqslant 
      \let\geq=\geqslant 
    \fi
  \fi
\fi 

\ifnfssone
  \newmathalphabet{\mathit}
  \addtoversion{normal}{\mathit}{cmr}{m}{it}
  \addtoversion{bold}{\mathit}{cmr}{bx}{it}
  \newmathalphabet{\mathbfit} 
  \addtoversion{normal}{\mathbfit}{cmr}{bx}{it}
  \addtoversion{bold}{\mathbfit}{cmr}{bx}{it}
  \newmathalphabet{\mathbfss} 
  \addtoversion{normal}{\mathbfss}{cmss}{bx}{n}
  \addtoversion{bold}{\mathbfss}{cmss}{bx}{n}
  \ifAMStwofonts
    \ifCUPmtlplainloaded \else
      \UseAMStwoboldmath
      \makeatletter
      \new@mathgroup\upmath@group
      \define@mathgroup\mv@normal\upmath@group{eur}{m}{n}
      \define@mathgroup\mv@bold\upmath@group{eur}{b}{n}
      \edef\UPM{\hexnumber\upmath@group}
      \new@mathgroup\amsa@group
      \define@mathgroup\mv@normal\amsa@group{msa}{m}{n}
      \define@mathgroup\mv@bold\amsa@group{msa}{m}{n}
      \edef\AMSa{\hexnumber\amsa@group}
      \makeatother
      \mathchardef\upi="0\UPM19
      \mathchardef\umu="0\UPM16
      \mathchardef\upartial="0\UPM40
      \mathchardef\leqslant="3\AMSa36
      \mathchardef\geqslant="3\AMSa3E

      \let\leq=\leqslant 
      \let\geq=\geqslant 
    \fi
  \fi
\fi 

\ifnfsstwo
  \DeclareMathAlphabet{\mathbfit}{OT1}{cmr}{bx}{it}
  \SetMathAlphabet\mathbfit{bold}{OT1}{cmr}{bx}{it}
  \DeclareMathAlphabet{\mathbfss}{OT1}{cmss}{bx}{n}
  \SetMathAlphabet\mathbfss{bold}{OT1}{cmss}{bx}{n}
  \ifAMStwofonts
    \ifCUPmtlplainloaded \else
      \DeclareSymbolFont{UPM}{U}{eur}{m}{n}
      \SetSymbolFont{UPM}{bold}{U}{eur}{b}{n}
      \DeclareSymbolFont{AMSa}{U}{msa}{m}{n}
      \DeclareMathSymbol{\upi}{0}{UPM}{"19}
      \DeclareMathSymbol{\umu}{0}{UPM}{"16}
      \DeclareMathSymbol{\upartial}{0}{UPM}{"40}
      \DeclareMathSymbol{\leqslant}{3}{AMSa}{"36}
      \DeclareMathSymbol{\geqslant}{3}{AMSa}{"3E}

      \let\leq=\leqslant 
      \let\geq=\geqslant 
    \fi
  \fi
\fi 

\ifCUPmtlplainloaded \else
  \ifAMStwofonts \else 
    \def\upi{\pi}
    \def\umu{\mu}
    \def\upartial{\partial}
  \fi
\fi

\title{Selection of quasar candidates from combined radio and optical surveys using neural networks}
\author[Carballo et al.]
{R. Carballo$^1$
\thanks{E-mail:carballor@unican.es},
 A.S. Cofi\~no$^1$, J.I. Gonz\'alez-Serrano$^2$ \\
        $^1$ Dpto. de Matem\'atica Aplicada y Ciencias de la Computaci\'on, Univ. de  Cantabria\\
       ETS Ingenieros de Caminos, Canales y Puertos, Avda de los Castros s/n
 39005 Santander, Spain\\
        $^2$ Instituto de F\'\i sica de Cantabria (CSIC-UC),  Avda de los Castros s/n
 39005 Santander, Spain}

\pagerange{\pageref{firstpage}--\pageref{lastpage}} \pubyear{2003}

\begin{document}

\maketitle

\label{firstpage}

\begin{abstract}

The application  of supervised  artificial neural networks  (ANNs)
for quasar  selection  from  combined   radio  and  optical
surveys  with photometric and morphological data  is investigated,
using the list of candidates and  their classification from White
et  al.  (2000). Seven input parameters and one output, evaluated
to 1 for quasars and 0 for nonquasars during the training,  were
used, with architectures 7:1  and  7:2:1.  Both models  were
trained on samples  of  $\sim 800$  sources and  yielded similar
performance on independent test samples,  with reliability as
large as 87\% at 80\% completeness (or 90 to 80\% for completeness
from 70 to 90\%). For comparison  the quasar fraction from the
original candidate list  was  56\%. The accuracy is similar to
that found by White et al. using supervised learning with oblique
decision trees and training samples of similar size. In view of
the large degree of overlapping between quasars and nonquasars in
the parameter space, this performance probably approaches the
maximum value achievable with this database.  Predictions of the
probabilities for  the 98 candidates without spectroscopic
classification in White et al. are presented and compared with the
results  from their work. The values obtained for the two ANN
models and  the decision trees are found to be in good agreement.
This  is the first analysis of the performance of ANNs for the
selection of quasars. Our work shows that ANNs provide a promising
technique for the selection of specific object types in
astronomical databases.

\end{abstract}


\section{Introduction}

In the recent  years large astronomical databases based  on surveys at
different  wavelengths  have  been  made  publicly  available  to  the
astronomical community. A full exploitation of these databases will be
only possible with the  help of artificial intelligence (hereafter AI)
tools,  which will allow  the selection,  classification and  even the
definition of particular object types within the databases.

Artificial  Neural   Networks  (ANNs  hereafter)  are   one  of
these tools. ANNs  have been applied  in astronomy for mainly  the
following problems: classification of stellar spectra (e.g.
Bailer-Jones, Irwin \& von Hippel 1998), morphological star/galaxy
separation (e.g. Bertin \&  Arnouts  1996), for morphological and
spectral classification of galaxies (Lahav et al. 1996; Folkes,
Lahav \& Maddox 1996; Firth, Lahav \& Somerville 2003) and, more
recently, for the estimation of photometric redshifts of galaxies
(Firth et al. 2003). A summary of the most relevant applications
of ANNs in astronomy can be found in Tagliaferri et al. (2003).

In this work  we investigate the application of ANNs  in a new
domain, which is  the effective selection  of quasar candidates.
Although at the end  an optical  spectrum will be  required to
confirm  the quasar classification and  determine its redshift, an
optimized selection of quasar  candidates  allows  to  obtain
large quasar  samples  with  a reduction of  telescope time. Large
quasar samples are  necessary to address important questions in
cosmology,  such as the  comparison of the space distribution of
quasars with that predicted by theory.

The test  is based  on a combined  radio and optical  survey including
photometric and  morphological data: the list of  quasar candidates in
White  et al.  (2000), drawn  from the  cross-correlation of  the Very
Large  Array FIRST Survey  and the  Automatic Plate  Measuring Machine
(APM) catalogue of the POSS-I photographic $E$ and $O$ plates (McMahon
\&   Irwin  1992).    White   et  al.    obtained  the   spectroscopic
classification of  1130 of the candidates, 636  (56\%) being confirmed
as quasars.  These quasars form  the FIRST Bright Quasar Survey of the
North Galactic Cap (FBQS-2 hereafter).  From their results the authors
explored, for  the first  time, the viability  of AI tools  to obtain,
from radio and optical photometric and morphological data, an a priori
selection of the best quasar  candidates. The used tool was supervised
learning with the oblique  decision tree classifier OC1 (Murthy, Kasif
\& Salzberg 1994).   The decision tree had a  single output per object
and the desired outputs or {\it targets} were set, during the learning
(training), to  1 for  quasars and to  0 for nonquasars.   The authors
found that a  decision tree classifier, trained on  data sets of about
800 objects, allowed to obtain  an efficient selection of the quasars,
producing  samples   with  reliability  as   high  as  80\%   at  90\%
completeness.
The  work by
White et al.  opened the idea  of the application of AI techniques for
the effective  selection of quasar candidates from  combined radio and
optical photometric  surveys. Following this idea, we  analysed - with
the same sample  of quasar candidates - a  different classifier, which
is supervised  learning with ANNs,  taking advantage of the  wealth of
training  algorithms included  in  the Matlab\footnote{MATLAB is a
trademark of The MathWorks, Inc.}
  Neural Network  Toolbox (http://www.mathworks.com/).

The layout of  the rest of the paper is as  follows.  Section 2 starts
with a brief  description of the list of  quasar candidates from White
et al., including the  selection criteria and the available parameters
from  FIRST  and APM.   Then  the  performance  of the  decision  tree
classifier reported  by White  et al. is  summarized. In Section  3 we
describe the  techniques applied for  the training and testing  of the
ANNs, we explore  the performance of the model,  basically in terms of
the  reliability   and  completeness  of  the  samples   of  the  best
candidates,  and  finally we  use  the  model  to predict  the  quasar
probabilities  for  the  98  FBQS-2 candidates  without  an  spectrum.
Through the  discussion, our results are compared  with those obtained
in White et al. The main conclusions are presented in Section 4.

\section[]{Quasar selection from the FBQS-2 \\*candidates via decision trees}

The candidates for  FBQS-2 (White et al. 2000)  were obtained from
the correlation of the  VLA FIRST radio survey (down  to $S_{\rm
1.4 ~GHz} \sim 1$  mJy) with  stellar sources on POSS-I(APM) with
$E  \leq 17.8$ and a blue  colour  ($O-E \leq 2$).
The spectroscopic classification  of 1130  of the 1238  candidates
yielded 636 quasars, 96 narrow line AGNs, 68 BL Lac objects, 190
HII galaxies, 52 passive galaxies and 88 stars. The selection
efficiency for quasars was  therefore 636/1130=56\% (704/1130=62\%
combining quasars  and BL Lac). White et al.  classified any
object with broad emission lines as quasar, i.e., they did not use
the conventional cut at $M_B = -23$ to exclude lower luminosity
objects. Fifty of the 636  quasars fall into this low luminosity
category.  The redshift range for the whole quasar sample was $z
\sim 0$ to $z \sim 3.5$

White  et al.  present diagrams  showing  that the
fraction of quasars varied with optical  magnitude, optical
colour, radio flux and radio-optical position  separation. Based
on these results, the authors  suggested  that  AI  methods  could
be  used to assign the candidates an a priori probability of being
quasars, $p$(Q), before taking  the spectra.  They  analysed the
performance  of the  oblique decision tree classifier OC1,
improved  by using 10 trees instead of a single one, with  a
weighted voting scheme. The  sample of quasars with spectroscopic
classification was  divided into  five
sets.  Setting aside  the first set, the remaining four were used
for the training and  the first one for the test.  Repeating the
procedure for  each  of the  sets,  the authors
could use all the objects for the training and all  the objects
for the test.


The  performance  of any  classifier  can  be  quantified through
two important parameters, which are the efficiency and the
completeness of the subsamples  built from  the classifier as  a
function of  the used threshold $p_C$(Q).  For this  case, the
efficiency (or reliability) is the number  of candidates above
$p_C$(Q) that are quasars  divided by the total number of
candidates above this threshold. The completeness is the number of
quasars that are included  above $p_C$(Q) divided by the  total
number of  quasars.  By  decreasing the  threshold $p_C$(Q) above
which candidates are accepted, the completeness of the sample is
increased, but  probably at  the cost of  efficiency. For  the
extreme case of  $p ({\rm Q}) \geq 0$ the FBQS-2  sample would be
complete, i.e. it would include all the quasars  among the
candidates, but the reliability would drop to  56\%.  White et al.
found  the voting decision  trees  to be  a successful classifier,
allowing to construct subsamples of candidates $\sim$ 87\%
reliable at completeness 70-80\% or  still very reliable, $\sim$
80\%,   at 90\%  completeness.
The authors used  seven input parameters: $E$, $O-E$, log$_{\rm
10}$  $S_p$ (where  $S_p$ is the FIRST peak flux density),
$S_i/S_p$  (where $S_i$  is the FIRST integrated flux density),
the radio-optical  separation, and the point spread functions
PSF($E$) and PSF($O$).


\section[]{Quasar selection from the FBQS-2 \\*candidates via ANNs}

\subsection{Fitting and testing technique}

\begin{figure*}
\psfig{figure=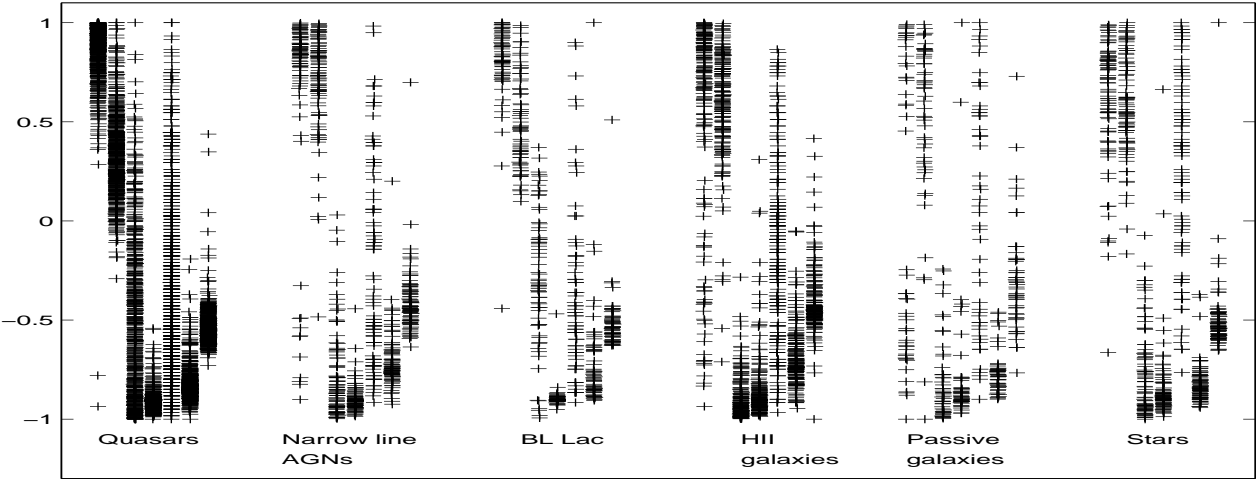,height=6.6cm,width=17cm}
 \caption{
Normalized parameters $E$, $O-E$, log$_{\rm 10}$ $S_{p}$,
$S_{i}/S_{p}$, radio-optical separation, PSF($E$) and PSF($O$),
ordered from left to right, for each class.}
\end{figure*}

An  ANN   is  a  computational   tool  which  provides   a
non-linear parametrized mapping between a set of input parameters
and one or more outputs. The type  of ANN we used is  the {\it
multi-layer perceptron} (hereafter MLP; Bishop 1995; Bailer-Jones,
Gupta \& Singh 2001). A particular ANN architecture may be denoted
as $N_{\rm in}: N_1:N_2:\ldots:N_{\rm out}$, where $N_{\rm in}$ is
the number of input  parameters, $N_{\rm 1}$ is  the number of
nodes in the first hidden layer  and so forth, and $N_{\rm out}$
is the number of nodes in  the output layer.   The nodes  are
connected, each connection  carrying a  {\it weight} and  each
node carrying a {\it bias}. For our work we assume that every node
is connected to  every node in the previous layer and every node
in the  next layer  only. Each node receives the output of all the
nodes in the previous layer and produces its own output, which
then feeds the nodes in the next layer. At a node in layer $s$ the
following calculation is obtained

\begin{equation}
z = \sum_{j=1}^m x_j^{s-1} w_{jk}^s + b_k,
\end{equation}
where $x_j^{s-1}$ are the inputs from the previous layer (with $m$
nodes), and $w_{jk}$ and $b_k$ are respectively the weights and
the bias of the node. Then the signal output of the node is $x_k^s
= g(z)$, where $g$ is the non-linear activation or transfer
function.  In order to obtain the correct mapping, a set of
representative input-output data is used for the training, a
process in which the weights and biases are optimized to minimize
the error in the outputs.

We used the same set of seven input parameters adopted by White et
al. (2000), normalizing each of them to the range $[-1,1]$.
In  order to  have homogeneous data sets we did not include the 28
candidates
missed in APM $E$  or $O$ or  both, and for which  White et  al.
used  the APS magnitudes (Minnesota Automated Plate Scanner POSS-I
catalogue, Pennington  et al. 1993). The  number  of FBQS-2
candidates  with spectral classification and homogeneous  data was
reduced to 1112 for this reason.  The distribution is as follows:
627 quasars, 94 narrow line AGNs, 67  BL Lac, 187  HII galaxies,
51 passive galaxies and 86 stars.

The  network output  consisted of  a  single node,  with the
target values  set  during   the  training  as  1  for   quasars
and  0 for nonquasars. For the output node we used a logarithmic
sigmoid activation function of the form $g(z) = 1 /[1 + exp(-z)]$,
in the range [0,1]. The actual output of the ANN  is then an
estimate of the probability that the source is a quasar, and it is
denoted as $p$(Q) (Richard and Lippmann 1991). The transfer
function for the hidden nodes was $g(z) = tanh (z) = 2 /[1 +
exp(-2z)]-1$, in the range $[-1,1]$.

Fig. 1 shows the normalized input parameters for the six classes.
Although the covered parameter space varies between classes, there
is a large degree of overlapping between quasars and nonquasars,
which will certainly limit the performance of the classification.
In fact, the category of nonquasars has an increased intrinsic
scatter due to the presence of classes with different physical
nature and covering different regions of the parameter space.

The error function we used was the mean of the squared errors, of
the form

\begin{equation}
mse = {1 \over N} \sum_{i=1}^N (p_i -t_i)^2 ,
\end{equation}
where $p_i$ and $t_i$  are, respectively, output (probability of
being a  quasar)  and target  value  for  the  $i$th object. The
sum of the squared errors has been widely used as the minimizing
error function for classification with the MLP (Richard and
Lippmann 1991, Bishop 1995, Lahav et al. 1996, Bailer-Jones et al.
2001, Ball et al. 2004). Although on theoretical grounds there are
more appropriate error functions for classification,
 such as {\it cross-entropy} (which assumes the
expected noise distribution for discrete variables), the
sum-of-squares error has proven to yield the same performance as
cross-entropy for MLP classification on real-world problems with
large databases  (Richard and Lippmann 1991). In addition, the
sum-of-squares error has the advantage that the determination of
the network parameters represents a linear optimization problem,
in particular, the powerful Levenberg-Marquardt algorithm for
parameter optimization is applicable specifically to a
sum-of-squares error function (Bishop 1995). Based on these
results, we used the $mse$ error function and applied the
Levenberg-Marquardt algorithm, which is the default optimization
technique used for {\it batch-training} (weights and biases
updated after all the input vectors are presented to the network)
in the Matlab Neural Network Toolbox. The Levenberg-Marquardt
algorithm is the fastest method for training moderate-sized neural
networks (Hagan \& Menhaj 1994).

Regardless of the optimization algorithm employed, one  of  the
main problems in the training process is that  of ``overfitting",
i.e. the ANN tends to memorize  the outputs, instead of modelling
the general intrinsic relationships in the data.  In order to
reduce   this  problem we used training with validation error.
With this  method, the training that is being carried out in the
\textit{training set} is automatically stopped when the error
obtained   running the trained   network in another  set, the
\textit{validation set}, does  not  decrease for a  given number
of iterations. We adopted for  this parameter, known as
\verb"maxfail" in Matlab, a value of 20, instead of the default
value of 5 iterations. An additional independent set, the \textit{
test set}, is  needed to evaluate the  ANN performance.


Following  the procedure  adopted  by  White et  al.,  we divided
the initial sample of candidates in  four subsets or folds (they
used five) of approximately similar sizes. Alike White et al., who
selected the folds  randomly, we chose them  to have similar
fractions of the different  object types  as the  total sample.
Setting aside each subset, the remaining three were used for the
training and validation, and the  subset itself was used  for the
test. The size of the test fold, of about 275 objects (i.e. 1/4 of
the candidates), was selected to insure the inclusion of about a
dozen of objects of the classes with fewer members, like passive
galaxies and  BL Lac. The three subsets used for training and
validation were firstly combined and then randomly divided in two
groups: one forming the training set, with 2/3 of the candidates,
and the other forming the validation set, with the remaining 1/3,
each of them with similar proportions of object types as the total
sample.
 Repeating the procedure for each of the four  folds, we obtained
four different classifiers, with the advantage  of having   used
all the objects for the training/validation and all the objects
for the test,  and therefore having optimized the statistics.

The  ANN was  run $10  \times  10$ times  per fold.  The first  factor
accounts  for the  obliged repetition  in an  algorithm  that includes
randomization (for example in the seeds for the initial weights of the
ANN) to avoid poor local minima. The second one arises from the use of
10 different  splittings to separate  the training and  the validation
sets.  In order  to choose  the best  ANN of  the 100  runs,  we first
selected the splitting  of the training and validation  sets that gave
the lowest value of the mean squared error, $\overline{mse}$, averaged
over the  10 fits. Adopting $mse  = 0.5 \times{mse}_{\rm  train} + 0.5
\times {mse}_{\rm valid}$ for each run,

\begin{equation}
\overline{mse} = 0.5 \times \overline{mse}_{\rm train} + 0.5
\times  \overline{mse}_{\rm valid}
\end{equation}
We then checked if the relation

\begin{equation}
\overline{|{mse}_{\rm train}-{mse}_{\rm valid}| / {mse}_{\rm train} } < 0.15
\end{equation} was satisfied for the splitting, to ensure that the errors in
the validation and
training were not only small, but also roughly similar (within
15\% on average). In the case that this condition failed, the
splitting with the next minimum value $\overline{mse}$ was checked
for condition (4) and so forth. Once the splitting was selected we
chose amongst the 10 ANNs the one with the minimum value of
$mse=0.5 \times {mse}_{\rm train} + 0.5 \times {mse}_{\rm valid}$
and satisfying

\begin{equation}
|{mse}_{\rm train}-{mse}_{\rm valid}| /
{mse}_{\rm train} < 0.15
\end{equation}

For a few  cases two or more  fits had the same minimum,  we took then
the first fit in running order. In the end we had a final ANN for each
of the  four test sets.  Running  each ANN for  its corresponding test
set we were able to obtain the values $p$(Q) for the 1112 candidates.

\subsection{Results}

We used  two different  ANN architectures. The  first one, denoted
as 7:1, does  not include hidden  layers and it  is also known as
a {\it logistic  discrimination model}.  The second architecture
includes a hidden layer  with two  nodes, and  it is denoted  as
7:2:1.   As we shall see, the performance of the classifier does
not improve with the inclusion of a hidden layer (increasing the
free parameters of the ANN from 8 to 19), therefore more complex
architectures were not explored. At the end of this subsection  we
present  the quasar probabilities obtained from the fitted ANNs
for  the list of FBQS-2 candidates without optical spectroscopy.


\subsubsection{Logistic discrimination model}

Fig.  2  shows $mse_{\rm  train}$ and $mse_{\rm  valid}$ for  the
400 networks  run (100 networks  per test  set $\times$  4 test
sets). We recall that each  group of 100 networks is divided  in
10 blocks, each of   them   corresponding   to    a   different
splitting   of   the validation-training sets, and  each block is
made of  10 fits. Some of the networks  or whole splittings
(blocks) produce peaks  in $mse_{\rm train}$, in $mse_{\rm valid}$
or in both.  The splittings showing peaks tend to have a  higher
$\overline{mse}$  than the splittings  lacking thereof, therefore
our choice of  the minimum  $\overline{mse}$ to  select the best
splitting. For 54\% of the networks the validation set stopped the
training.
The number of iterations  for the cases of validation stop is very
small (below 12), with an average of $\sim$ 4 compared to the
average $\sim$ 23 found for the networks stopped because of  other
reasons.


\begin{figure}
\psfig{figure=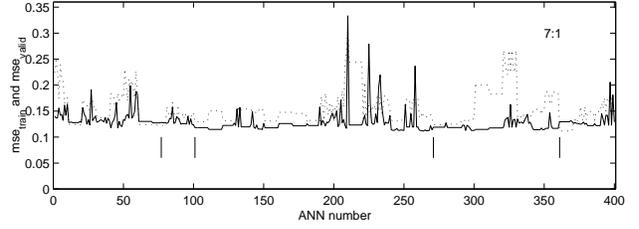,height=3.1cm}
 \caption{
 $mse_{\rm train}$ (continuous line) and $mse_{\rm
valid}$ (dotted line) for the 400 ANNs run. The vertical lines
mark the best ANN for each of the test sets.}
\end{figure}

\begin{figure}
\psfig{figure=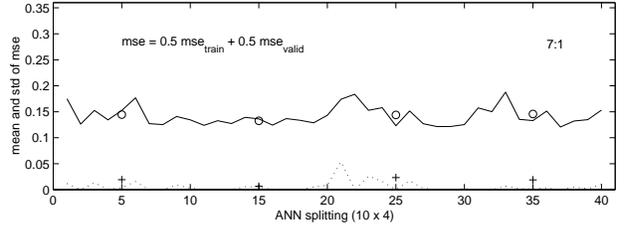,height=3.1cm} \caption{Mean (continuous
line) and standard deviation (dotted line) of $mse$ ( $= 0.5
\times mse_{\rm train} + 0.5 \times mse_{\rm valid}$) over the 10
runs with similar {\it test}, {\it training} and {\it validation}
sets. The first parameter is denoted as $\overline{mse}$ in the
text. The meaning of the circles and crosses  is explained in the
text.
}
\end{figure}

Fig. 3 shows $\overline{mse}$ and the standard deviation of $mse$
over the ten  runs for each of  the 40 different  splittings (10
splittings per test set $\times$ 4 test sets). The scatter ranges
from 0 to about 0.05, with an average for the 40 splittings of
0.006. The influence of the  initial values  on the  performance
of the  selected network  is negligible, since the standard
deviation of $mse$ due to different initiations is clearly  much
lower than the  mean values. The same  occurs for the separation
of the training and validation sets; the circles in Fig.  2 show
the average of  $\overline{mse}$ over the 10 different splittings
per test  set, and  the averages are  significantly larger  than
their standard deviations, symbolized as  crosses.  Finally, the
figure also shows that the  four mean values (one per test  set)
are very similar, with the  standard deviation being  much lower
than the average (this average and standard deviation are  not
shown in the figure). The last result  demonstrates that  the
performance of the  network does  not depend strongly on the
particular selected {\it test set} either. The 4 best  ANNs (one
per test set) are marked  with vertical lines in Fig. 2 and Table
1 summarizes some of their parameters.

\begin{table}
 \centering
 \begin{minipage}{80mm}
  \caption{Parameters of the 4 selected ANNs for the logistic model}
  \begin{tabular}{ccccc}
\hline
$mse_{\rm train}$     & $mse_{\rm valid}$         & $N_{\rm iter}$    &$mse_{\rm test}$&$\overline E$ \\
                  &                       &             &\\
\hline
0.128 & 0.122 &  ~~~3 *  &0.119 &0.49\\
0.118 & 0.130 & 18       &0.133 &0.54\\
0.119 & 0.124 & 23       &0.154 &0.63\\
0.129 & 0.112 & ~~~11 *  &0.122 &0.49\\
\hline
\end{tabular}

* :  The  results on the validation set stopped the
training.

\end{minipage}
\end{table}

So  far we  have discussed  the mean  squared errors  obtained  in
the training process. Column four of Table 1 gives the mean
squared errors obtained  for  the  {\it  test  sets}. The  latter
{\it mse}  values generally show a  good agreement with those
obtained for the training and validation sets. However, a more
interesting parameter for  the purpose of assessing  the
performance  of  the network  is  the normalized  error function
(Bishop 1995), of the form

\begin{equation}
\overline E = \sum_{i=1}^N (p_i - t_i)^2 / \sum_{i=1}^{N}
(\overline t - t_i)^2,
\end{equation}
where  $\overline t$  is the  mean of  the target  data over  the
test set. This  error function  equals unity  when the model  is
as  good a predictor of the target data as the simple model $p=
\overline t$, and equals zero if  the model predicts the data
values exactly. The value we found  is around 0.55.   Although the
model  is not good  enough for classification, the results are
powerful  for our aim of selecting the best  candidates.  In fact,
compared  to  the  model that  takes  $p= \overline t$, which
would  give $mse_{\rm test} \sim 0.246$,  the $mse$ obtained with
the ANNs is reduced about a factor two. In the next paragraphs we
present the results  of  the model  in terms  of completeness and
efficiency of  the  subsamples of  {\it  the best candidates} that
can be drawn from the ANN model.

Fig. 4  shows the distribution of  $p$(Q) for the  1112 candidates
and the  logistic  discrimination  ANN. The  model gives
probabilities above 0.5 for most of the quasars, although there is
a large number of them  with  probabilities below  this  value.
Narrow  line AGNs,  HII galaxies, passive galaxies and stars tend
to  give low probabilities, and the  model provides  therefore a
good  means to reject  objects of these types. The  number of  BL
Lac  objects is  small, and  their distribution of $p$(Q) is
rather   flat, and even slightly increased at high probabilities,
therefore the current model is not able to reject these sources.
The last problem was also found by White et al. using the decision
tree classifier.

\begin{figure}
\psfig{figure=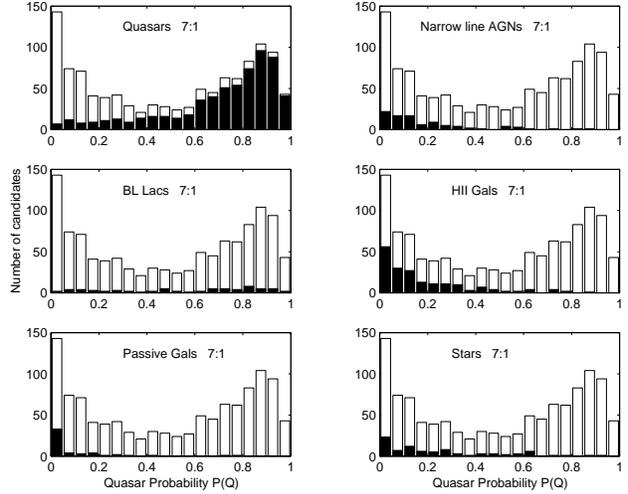,height=6.6cm} \caption{Distribution of
$p$(Q) for the logistic discrimination model. The shaded
distributions correspond to the objects of the indicated types
(quasars, narrow line AGNs, BL Lacs, etc).}
\end{figure}

\begin{figure}
\psfig{figure=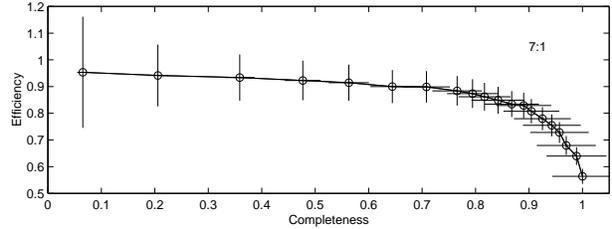,height=3.1cm }
  \caption{Efficiency vs. completeness of the sample for
$p$(Q) $> 0,0.05,0.1,0.15,0.2, \ldots 0.95$ (right to left) and
the logistic model. Poissonian errors were assumed.}
\end{figure}

\begin{figure}
\psfig{figure=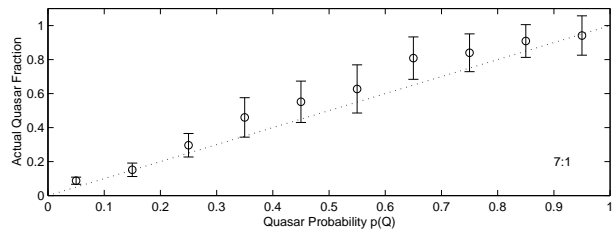,height=3.1cm }
 \caption{Fraction of candidates that are quasars as a function of
$p$(Q) for the logistic model. Poissonian errors were assumed.}
\end{figure}

The efficiency  and completeness  of the sample  as a function  of
the quasar  probability  threshold  $p_C$(Q)  are  shown in  Fig.
5.  The logistic  model  allows  to  obtain  a  high  reliability
at  a  high completeness: for completenesses of  70, 80 and 90\%
the corresponding reliabilities  are 90,  87  and 81\%.  Fig.  6
shows  the fraction  of candidates that are quasars as  a function
of $p$(Q).  The fraction is slightly above $p$(Q) (about 0.1 in
the range from $p$(Q) 0.3 to 0.8), i.e.  the likelihood  that a
candidate turns  out to  be a  quasar is slightly larger than the
probability given by the model.

Fig. 4 shows that the  majority of the high-$p$(Q) candidates that
are not quasars  are BL  Lac objects. Taking  $p$(Q)$>$0.75 there
are 353 quasars,  24 BL  Lacs, two  narrow line  AGNs, three  HII
galaxies, a passive galaxy and three stars. An inspection of the
input parameters for the nonquasars revealed as the most
outstanding result that the {\it whole population} of BL Lacs has
radio fluxes higher than those found for the remaining nonquasar
classes, and similar to those typically found in quasars (see Fig.
7a). About 36\% of the BL Lac have $p$(Q)$>$0.75 and Figs. 7b and
7c show that these correspond to the cases with bluer $O-E$
colours. The efficiency  of quasar selection using the cut at
$p$(Q)$=$0.75 is 91\% (353/386) and increases to 98\% considering
quasar or BL Lac selection (377/386).  The corresponding
completeness  would be 56\% (353/627) for quasars and 54\%
(377/694) for  quasars or  BL Lac. The completeness decreases in
the latter case since {\it only} blue BL Lac are confused with
quasars.

\begin{figure}
\psfig{figure=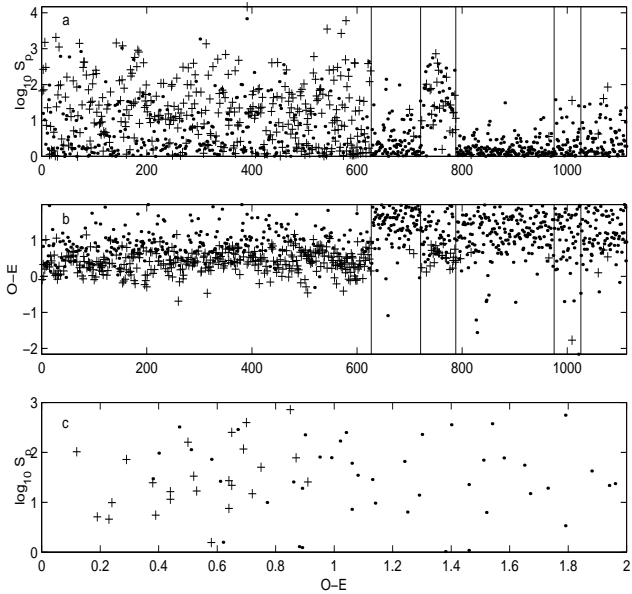,height=8.0cm,width=8.4cm} \caption{{\bf
a} and {\bf b}: Log$_{10}$$(S_p)$ and $O-E$ for the classes of
quasars, narrow line AGNs,  BL Lac, HII galaxies, passive galaxies
and stars (separated by vertical lines). {\bf c}:
Log$_{10}$$(S_p)$ versus $O-E$ for the BL Lac. Crosses correspond
to $p$(Q)$>$0.75 for the 7:1 model.}
\end{figure}

\begin{figure}
\psfig{figure=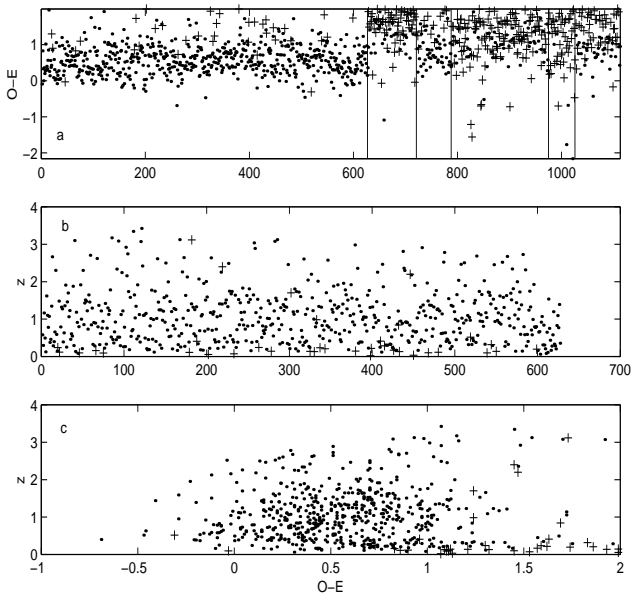,height=8.0cm,width=8.4cm} \caption{{\bf
a}: $O-E$ for the classes of quasars, narrow line AGNs, BL Lac,
HII galaxies, passive galaxies and stars (separated by vertical
lines). {\bf b} and {\bf c}: $z$ and  $z$ versus $O-E$ for the
quasars. Crosses correspond to $p$(Q)$<$0.2 for the 7:1 model.}
\end{figure}

At  the  other  extreme,  there  are  36  quasars  with
probabilities $p$(Q)$<$0.2, and their most significant differences
with respect to the remaining quasars are their redder $O-E$
colours and lower redshifts, with twenty-five of them at $z<0.25$
(see Figs. 8a, 8b, 8c). The misclassified quasars also differ,
although to a lower extent, in their larger integrated-to-peak
radio flux ratio, larger optical-radio separation and wider PSF.
The low probabilities found for the low-$z$ quasars should not be
regarded  as a limitation of the classifier, since at low
redshifts the host galaxy is expected to be slightly resolved and
to have a noticeable contribution to the total ''galaxy + quasar''
emission. This contribution, imperceptible at higher redshifts, is
the most likely explanation for the differences in the input
parameters between low-$z$ quasars and the remaining quasars.

If only the quasars with $z \geq 0.25$ are considered, the
fraction of them with $p$(Q)$<$0.2 drops from 6\% (36/627) to 2\%
(11/558). As for the probability cut $p$(Q)$>$0.75, the
 efficiency remains at 91\%  and the completeness increases from 56\%
 to 62\%.  White et al. also found, using the decision tree classifier, that
 the  great majority of quasars with low probabilities were at low
 redshift (out of 30 quasars with $p(Q)<0.2$, 24 had $z <
 0.25$).

\begin{figure}
\psfig{figure=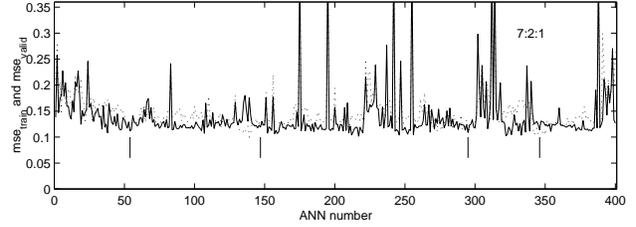,height=3.1cm}
 \caption{Similar to Fig. 1, but for the 7:2:1 architecture.}
\end{figure}

\begin{figure}
\psfig{figure=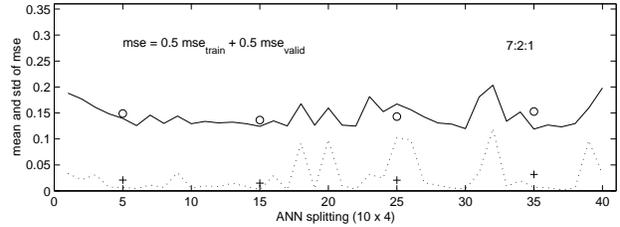,height=3.1cm} \caption{Similar to Fig.
2, but for the 7:2:1 architecture.}
\end{figure}

\subsubsection{ANN with a hidden layer}

In this  subsection we  present the results  for the ANN  model
7:2:1. Fig. 9 shows $mse_{\rm train}$ and  $mse_{\rm valid}$ for
the 400 networks. Three main differences with respect to the model
without a hidden layer are revealed: (i) the distribution of
$mse_{\rm train}$ is more noisy but there is a better agreement
between $mse_{\rm train}$ and $mse_{\rm valid}$, (ii) validation
stopping dominates (96\% of the cases) over the remaining  reasons
to stop the training and (iii) the number of iterations  for  the
cases of validation  stopping reaches higher values, with an
average $\sim$ 33 iterations. Fig. 10 shows the average and
standard deviation of $mse$ over the ten runs for each of  the 10
splittings and for each of  the 4 test sets. The large variations
of $mse$ are clearly evident  from this figure: the standard
deviation of $mse$ ranges from 0.002 to 0.12, with an average for
the 40 splittings of 0.027, i.e.    4.5 times larger than   for
the   7:1 architecture. However, for most of the
training-validation-test configurations the scatter of $mse$ is
still much lower than the mean value. Considering the averages per
test set,  the mean values for $\overline{mse}$ are also
significantly larger  than their standard deviations (denoted with
circles and  crosses respectively)  and the same occurs
considering the average  over the four test sets. As occurred  for
the 7:1 model,  the performance  of the selected network does  not
depend strongly on changes of the initiation values, splitting for
training-validation or choice of test set.

\begin{table}
 \centering
 \begin{minipage}{80mm}
  \caption{Parameters of the 4 selected ANNs for the model 7:2:1}
  \begin{tabular}{ccccc}
\hline
$mse_{\rm train}$& $mse_{\rm valid}$& $N_{\rm iter}$&$mse_{\rm test}$&$\overline E$ \\
                 &                  &               &\\
\hline
0.112 & 0.126 & 14 *  &0.110 &0.45\\
0.119 & 0.121 & 8     &0.127 &0.52\\
0.109 & 0.118 & 16 *  &0.146 &0.60\\
0.114 & 0.106 & 27 *  &0.127 &0.51\\
\hline
\end{tabular}
\end{minipage}
\end{table}

The relevant parameters of the 4 selected ANNs are summarized in
Table 2. The $mse$ values for  the test sets  generally show  a
good agreement with  the values obtained for the train and
validation sets. Both $mse_{\rm test}$ and the normalized error
function $\overline  E$ are on average similar to those obtained
for the 7:1 architecture.

\begin{figure}
\psfig{figure=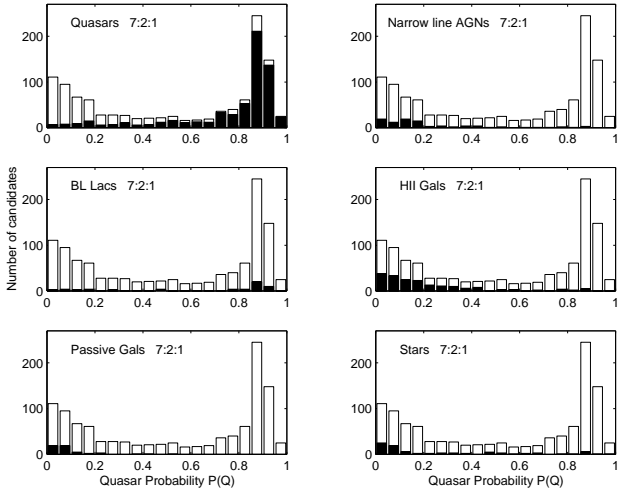,height=6.6cm} \caption{Similar to Fig.
3, but for the 7:2:1 architecture.}
\end{figure}

Fig. 11  shows the distribution of  $p$(Q) for the  1112
candidates and the 7:2:1 model.  The distribution  is more peaked
towards the extreme values  of the  probabilities than  in  the
logistic model.  In  this respect, the  7:2:1 model  gives a
better agreement with  the results from  OC1 than  the logistic
model. As occurred for the logistic model and OC1, all the
nonquasar classes except for the BL Lac tend to give low
probabilities.

The efficiency  and completeness  of the sample  as a function  of
the quasar probability threshold are very similar to the values
found for the logistic model. For completenesses of 70, 80 and
90\% the corresponding reliabilities are 88, 87 and 81\%
respectively. Again there is a very  good agreement between $p$(Q)
and the likelihood that a candidate with $p$(Q) turns out to be a
quasar (measured by the fraction of candidates at this $p$(Q) that
are quasars), except for $p$(Q) around 0.55, where the likelihood
is increased by an amount around 0.1.

\begin{table*}
 \centering
 \begin{minipage}{160mm}
  \caption{Efficiency and completeness of quasar
  selection from the sample of FBQS-2 candidates using ANNs}
\begin{tabular}{llrcllr}
\hline
\multicolumn{7}{l}{Size of spectroscopically identified sample 1112 }\\
\multicolumn{7}{l}{Training+validation set size  $\sim$ 840 }\\
\multicolumn{7}{l}{Total number of quasars ~627}\\
\hline
\multicolumn{3}{l}{ANN 7:1}&&\multicolumn{3}{l}{ANN 7:2:1 }\\
\multicolumn{3}{l}{Completeness/Efficiency ~~70\% / 90\%}&&
\multicolumn{3}{l}{Completeness/Efficiency ~~70\% / 88\%}\\
   \multicolumn{3}{l}{~~~~~~~~~~~~~~~~~~~~~~~~~~~~~~~~~~~80\% / 87\%} &&
   \multicolumn{3}{l}{~~~~~~~~~~~~~~~~~~~~~~~~~~~~~~~~~~~80\% / 87\%}\\
   \multicolumn{3}{l}{~~~~~~~~~~~~~~~~~~~~~~~~~~~~~~~~~~~90\% / 81\%} &&
   \multicolumn{3}{l}{~~~~~~~~~~~~~~~~~~~~~~~~~~~~~~~~~~~90\% / 81\%}\\
  $p$(Q)$<$0.2&Quasars     &36  &     & $p$(Q)$<$0.2 & Quasars    &39     \\
   &&                           &                       &&& \\
 $p$(Q)$>$0.75&Candidates  & 386&     &$p$(Q)$>$0.85 & Candidates & 418   \\
            &Quasars       & 353&                   && Quasars    & 372   \\
            &BL Lac        &  24&                   && BL Lac     &  30   \\
&Efficiency for quasars&91\%&         && Efficiency for quasars&89\%  \\
&Efficiency for quasars + BL Lac&98\%&&&Efficiency for quasars + BL Lac&96\%\\
&Completeness for quasars&56\%          &&&Completeness for quasars&59\%  \\
&Completeness for quasars + BL Lac&54\%&&&Completeness for quasars + BL Lac&58\%\\
\hline
\end{tabular}
\end{minipage}
\end{table*}

Taking $p_C$(Q)=0.85, there are 372 quasars, 30 BL Lacs, 3 narrow
line AGNs, five  HII galaxies and eight stars above this cut. A
result similar to the one obtained in Fig. 7 for the logistic
model is found for the 7:2:1 architecture: the majority of the
high-$p$(Q) nonquasars are blue BL Lac objects. The efficiency of
quasar selection  for this threshold is  89\% and increases to
96\% considering quasar or BL  Lac selection.  The corresponding
completeness would be 59\% for  quasars and 58\% for quasars or BL
Lac.

Regarding the limit of low probabilities we find thirty-nine
quasars with $p$(Q)$<$0.2, twenty-two of them with redshifts below
0.25. We find similar results to those presented in Fig. 8 for the
logistic model: most of the misclassified quasars have lower
redshifts and redder $O-E$  colours than the remaining quasars, as
well as higher integrated-to-peak radio flux ratios and wider
PSFs, and these results are indicative of an appreciable
contribution of the emission from the host galaxy.  If only the
quasars with $z \geq 0.25$ are considered, the fraction of them
with $p$(Q)$<$0.2 decreases from 6\% to 3\%. As for the
probability cut $p$(Q)$>$0.85, the
 efficiency would remain at 89\%  and the completeness would increase from 59\%
 to 65\%.

Table 3 presents a summary of the performance of the two ANN
models.  Both use similar training set  sizes and achieve similar
efficiencies  for completeness  in the range from 70 to 90\%. The
main difference is that the distribution of $p$(Q) is more peaked
towards  the extreme values  (0 and 1)  for the model with a
hidden layer than for the logistic one.

Fig. 12 shows that the distribution of efficiency versus
completeness for the two ANN models and the oblique decision tree
OC1 are very similar. The agreement obtained for the three
different classifiers favours  the interpretation that the found
accuracy - $\sim$ 87\% at 80\% completeness - is more limited by
the data structure itself (i.e. the large degree of  overlapping
between quasars and nonquasars in the input parameter space) than
by the complexity of the algorithms. The ANN and the decision tree
classifiers both point to  BL Lac (blue BL Lac for ANNs) and
low-$z$ quasars as the object types that most severely limit the
accuracy of quasar selection, the former producing intruders
(false alarms), and the latter misclassifications.

\begin{figure}
\psfig{figure=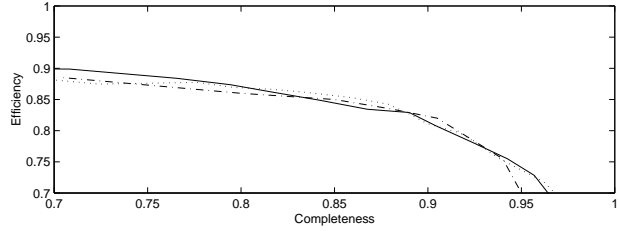,height=3.1cm }\caption{Efficiency versus
completeness for the 7:1 ANN (continuous line), the 7:2:1 ANN
(dotted line) and the OC1 decision tree (dash-dotted line).}
\end{figure}

Owens et al. (1996) apply the decision tree OC1 for the
morphological classification of galaxies taken from the ESO-LV
catalogue (Lauberts \& Valentijn 1989), and present a comparison
of their results with those obtained by Storrie-Lombardi et al.
(1992)  for the same sample using ANNs. The classification {\it
into six classes} has an overall efficiency around 63\% for the
two methods, with a difference between them lower than 3\%. Owens
et al. (1996) attribute the found similarity to limitations in the
classification accuracy intrinsic to the database (errors in the
assumed classification for some of the galaxies and a poorly
defined separation between classes). Our work and Owens et al.
(1996) show examples of classification from astronomical databases
in which OC1 and ANNs give similar performances, probably at the
limit set by the database itself, whose attributes do not provide
enough information for a more accurate classification.

\subsubsection{Predictions for the FBQS-2 candidates without spectroscopic
classification}

\begin{figure}
\psfig{figure=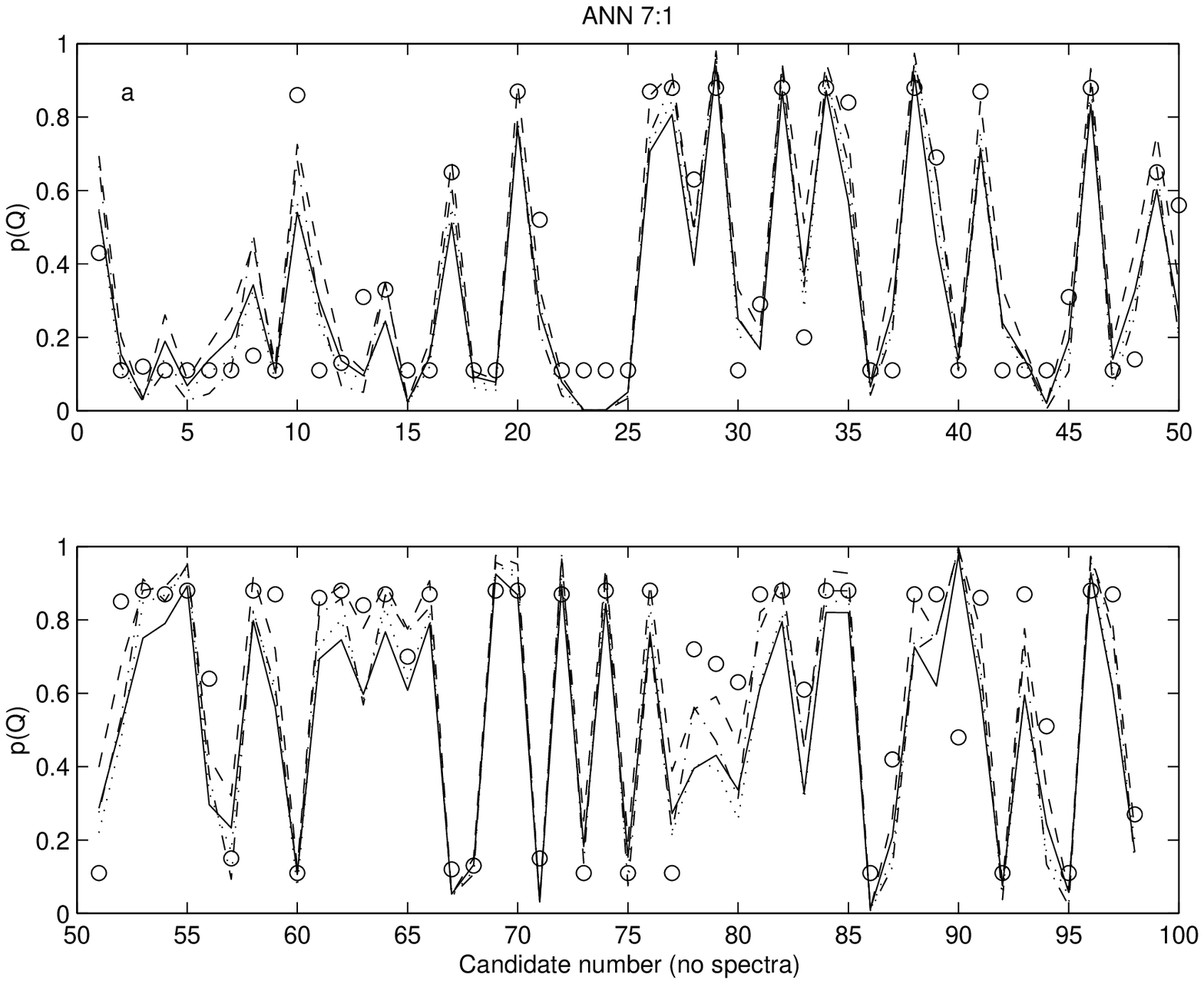,height=6.6cm }
\psfig{figure=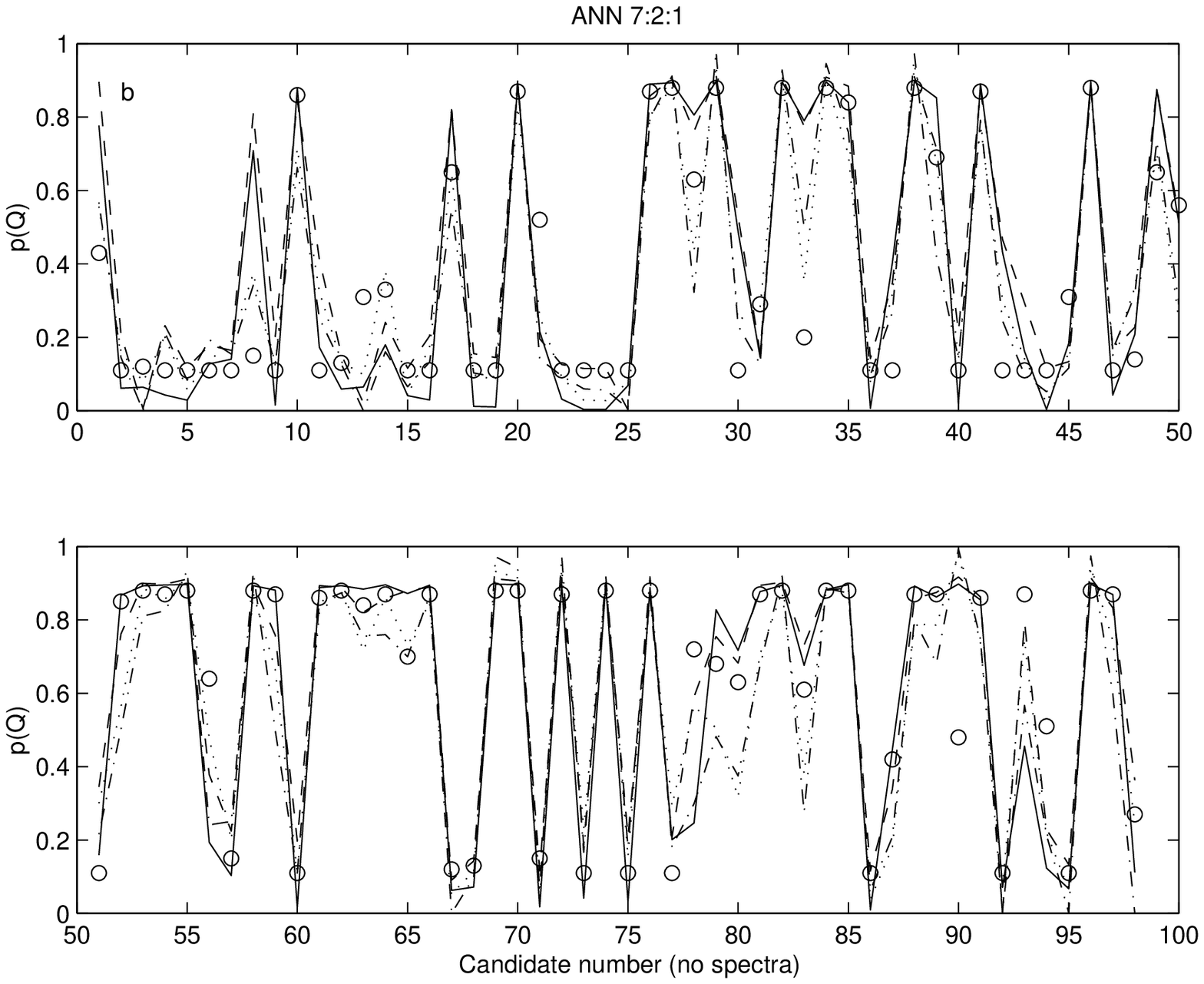,height=6.6cm } \caption{ $p$(Q)
distribution for the 98 FBQS-2 candidates without spectroscopic
classification in White et al. (2000) using the ANN architectures
7:1 ({\bf a}) and 7:2:1 ({\bf b}). The four line types correspond
to the four selected ANNs. Circles correspond to  the  $p$(Q)
values obtained by White et al. with OC1.}
\end{figure}

The ANN models  7:1 and 7:2:1 were used  to estimate the
probabilities $p$(Q) for the 98 FBQS-2 candidates without spectral
classification in White  et  al.   We  adopted  four classifiers
per model, corresponding to the four selected  ANNs (parameters
described in Tables 1 and 2).  Fig. 13a shows  the probabilities
obtained with the 7:1 model - plotted with a different line type
for each ANN -  and using OC1 (White  et al.). There is  a good
agreement between  the probabilities predicted with  the four ANNs
and between them and  the values from OC1. Similar results are
found for the 7:2:1 model (Fig. 13b). The probabilities obtained
for the ANN models and OC1 are listed in Table 4. For the ANN
models we give the mean and standard deviation of $p$(Q) over the
four selected ANNs.

\begin{table*}
 \centering
 \begin{minipage}{160mm}
\caption{Quasar probabilities for the 98 candidates without
spectroscopic classification in White et al. (2000)}
\begin{tabular}{|c|c|c|c|c|c|c|c|c|c|c|c|c|}
\hline Name& \multicolumn{2}{c}{ANN 7:1}   &
\multicolumn{2}{c}{ANN7:2:1}& OC1 && Name&
 \multicolumn{2}{c}{ANN 7:1}   &
 \multicolumn{2}{c}{ANN 7:2:1} & OC1\\
FBQS J &  $\overline p$(Q) & $\sigma$  &  $\overline p$(Q) &
$\sigma$  &$p$(Q) && FBQS J & $\overline p$(Q) & $\sigma$  &
  $\overline p$(Q) & $\sigma$  &$p$(Q)\\
\hline
071505.4+340501& 0.61& 0.08& 0.70& 0.17& 0.43&&122251.3+331640& 0.25& 0.07& 0.39& 0.15& 0.56\\
071650.6+350520& 0.15& 0.04& 0.14& 0.06& 0.11&&122407.3+375332& 0.30& 0.07& 0.26& 0.08& 0.11\\
071903.2+342550& 0.03& 0.01& 0.04& 0.04& 0.12&&122520.4+292420& 0.55& 0.09& 0.67& 0.17& 0.85\\
073018.1+224502& 0.17& 0.07& 0.17& 0.09& 0.11&&122856.6+355635& 0.85& 0.07& 0.87& 0.04& 0.88\\
073237.9+342952& 0.06& 0.02& 0.07& 0.04& 0.11&&123659.5+423641& 0.85& 0.04& 0.87& 0.03& 0.87\\
073317.3+223725& 0.12& 0.06& 0.16& 0.03& 0.11&&123757.9+223430& 0.94& 0.03& 0.91& 0.02& 0.88\\
073833.5+360957& 0.18& 0.07& 0.16& 0.02& 0.11&&124327.8+232811& 0.36& 0.06& 0.32& 0.13& 0.64\\
074342.2+321543& 0.40& 0.08& 0.56& 0.24& 0.15&&124444.5+223305& 0.20& 0.10& 0.19& 0.06& 0.15\\
082711.2+223323& 0.10& 0.02& 0.12& 0.08& 0.11&&124840.4+241240& 0.84& 0.05& 0.90& 0.02& 0.88\\
083522.7+424258& 0.63& 0.09& 0.78& 0.10& 0.86&&124958.8+245233& 0.62& 0.07& 0.70& 0.16& 0.87\\
085552.7+384325& 0.31& 0.08& 0.29& 0.10& 0.11&&125018.1+364914& 0.10& 0.02& 0.11& 0.07& 0.11\\
085624.8+345024& 0.12& 0.04& 0.12& 0.04& 0.13&&125142.2+240435& 0.77& 0.07& 0.86& 0.03& 0.86\\
091309.2+413635& 0.09& 0.03& 0.05& 0.04& 0.31&&125256.9+252503& 0.82& 0.06& 0.88& 0.01& 0.88\\
091833.8+315620& 0.30& 0.06& 0.24& 0.10& 0.33&&125444.7+425305& 0.63& 0.10& 0.79& 0.07& 0.84\\
091845.7+233833& 0.02& 0.00& 0.07& 0.04& 0.11&&131823.4+262623& 0.84& 0.06& 0.84& 0.06& 0.87\\
093456.7+263054& 0.15& 0.03& 0.14& 0.08& 0.11&&131848.3+252815& 0.70& 0.09& 0.80& 0.09& 0.70\\
101355.2+300546& 0.59& 0.07& 0.71& 0.13& 0.65&&132324.1+251809& 0.84& 0.05& 0.87& 0.02& 0.87\\
102802.9+304743& 0.08& 0.02& 0.09& 0.06& 0.11&&134531.0+255504& 0.05& 0.00& 0.05& 0.04& 0.12\\
102857.6+344054& 0.07& 0.01& 0.08& 0.06& 0.11&&134540.0+280123& 0.13& 0.02& 0.11& 0.03& 0.13\\
103346.3+233220& 0.81& 0.05& 0.88& 0.03& 0.87&&140819.3+294950& 0.95& 0.02& 0.93& 0.03& 0.88\\
103818.1+424442& 0.28& 0.05& 0.20& 0.05& 0.52&&141257.7+232618& 0.92& 0.04& 0.91& 0.02& 0.88\\
105330.9+331342& 0.07& 0.02& 0.09& 0.04& 0.11&&143655.7+234928& 0.03& 0.00& 0.06& 0.04& 0.15\\
105653.3+331945& 0.00& 0.00& 0.05& 0.05& 0.11&&144053.9+270642& 0.94& 0.04& 0.93& 0.03& 0.87\\
110113.8+323155& 0.00& 0.00& 0.05& 0.05& 0.11&&144755.7+382813& 0.19& 0.05& 0.16& 0.08& 0.11\\
112242.8+414355& 0.04& 0.01& 0.04& 0.04& 0.11&&145007.2+315050& 0.90& 0.05& 0.91& 0.01& 0.88\\
113020.4+422204& 0.76& 0.07& 0.84& 0.04& 0.87&&150228.5+354455& 0.13& 0.05& 0.14& 0.08& 0.11\\
113124.2+261951& 0.87& 0.05& 0.90& 0.01& 0.88&&150428.0+262419& 0.81& 0.07& 0.88& 0.03& 0.88\\
113324.7+323449& 0.45& 0.05& 0.58& 0.24& 0.63&&150435.8+335728& 0.27& 0.08& 0.22& 0.05& 0.11\\
113442.0+411330& 0.96& 0.02& 0.93& 0.04& 0.88&&150555.4+424415& 0.48& 0.09& 0.43& 0.18& 0.72\\
113609.0+360641& 0.26& 0.06& 0.37& 0.16& 0.11&&151314.9+342111& 0.47& 0.08& 0.64& 0.17& 0.68\\
113639.1+372651& 0.19& 0.03& 0.18& 0.07& 0.29&&151627.3+305220& 0.34& 0.09& 0.52& 0.21& 0.63\\
113707.7+290324& 0.92& 0.03& 0.91& 0.01& 0.88&&151913.4+252134& 0.71& 0.10& 0.79& 0.11& 0.87\\
113921.2+350748& 0.37& 0.10& 0.61& 0.21& 0.20&&152049.1+375219& 0.85& 0.06& 0.90& 0.02& 0.88\\
114048.0+332908& 0.91& 0.03& 0.91& 0.02& 0.88&&152158.4+381814& 0.39& 0.08& 0.52& 0.22& 0.61\\
114111.1+300442& 0.65& 0.08& 0.80& 0.08& 0.84&&152547.2+425210& 0.87& 0.05& 0.88& 0.00& 0.88\\
115244.4+311123& 0.06& 0.01& 0.08& 0.06& 0.11&&153402.2+425249& 0.87& 0.04& 0.88& 0.01& 0.88\\
115943.8+303348& 0.28& 0.07& 0.33& 0.06& 0.11&&153411.3+262124& 0.01& 0.00& 0.06& 0.05& 0.11\\
120354.7+371137& 0.95& 0.03& 0.93& 0.03& 0.88&&153420.2+413007& 0.19& 0.06& 0.29& 0.12& 0.42\\
120908.4+265131& 0.57& 0.09& 0.67& 0.18& 0.69&&153521.6+331826& 0.76& 0.07& 0.83& 0.05& 0.87\\
121147.1+240736& 0.13& 0.03& 0.13& 0.08& 0.11&&153818.6+410548& 0.73& 0.07& 0.80& 0.09& 0.87\\
121232.3+425821& 0.76& 0.06& 0.83& 0.06& 0.87&&154007.6+252836& 0.99& 0.01& 0.95& 0.05& 0.48\\
121355.3+365255& 0.26& 0.05& 0.36& 0.11& 0.11&&154049.2+390351& 0.67& 0.08& 0.80& 0.07& 0.86\\
121529.6+391200& 0.14& 0.03& 0.17& 0.08& 0.11&&155537.5+221327& 0.06& 0.02& 0.04& 0.04& 0.11\\
121727.8+290449& 0.02& 0.01& 0.05& 0.05& 0.11&&155723.9+420825& 0.67& 0.09& 0.64& 0.16& 0.87\\
121902.5+222416& 0.18& 0.06& 0.15& 0.03& 0.31&&160531.1+243147& 0.22& 0.09& 0.20& 0.05& 0.51\\
122004.3+311148& 0.89& 0.04& 0.89& 0.01& 0.88&&162237.8+235943& 0.05& 0.02& 0.07& 0.05& 0.11\\
122034.6+363357& 0.12& 0.06& 0.13& 0.06& 0.11&&163718.8+272607& 0.96& 0.02& 0.93& 0.03& 0.88\\
122208.1+240012& 0.33& 0.08& 0.27& 0.06& 0.14&&164733.9+364055& 0.69& 0.09& 0.76& 0.12& 0.87\\
122221.3+372335& 0.66& 0.06& 0.80& 0.09& 0.65&&170753.9+272418& 0.18& 0.04& 0.18& 0.16& 0.27\\
\hline
\end{tabular}
\end{minipage}
\end{table*}

\begin{figure}
\psfig{figure=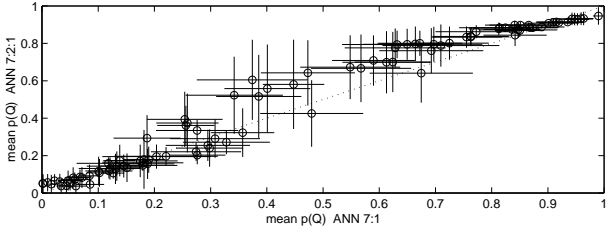,height=3.1cm } \caption{Comparison
between $\overline p$(Q) for the 7:1 and  the 7:2:1 ANNs, for the
FBQS-2 candidates without spectroscopic classification in White et
al. (2000).}
\end{figure}
\begin{figure}
\psfig{figure=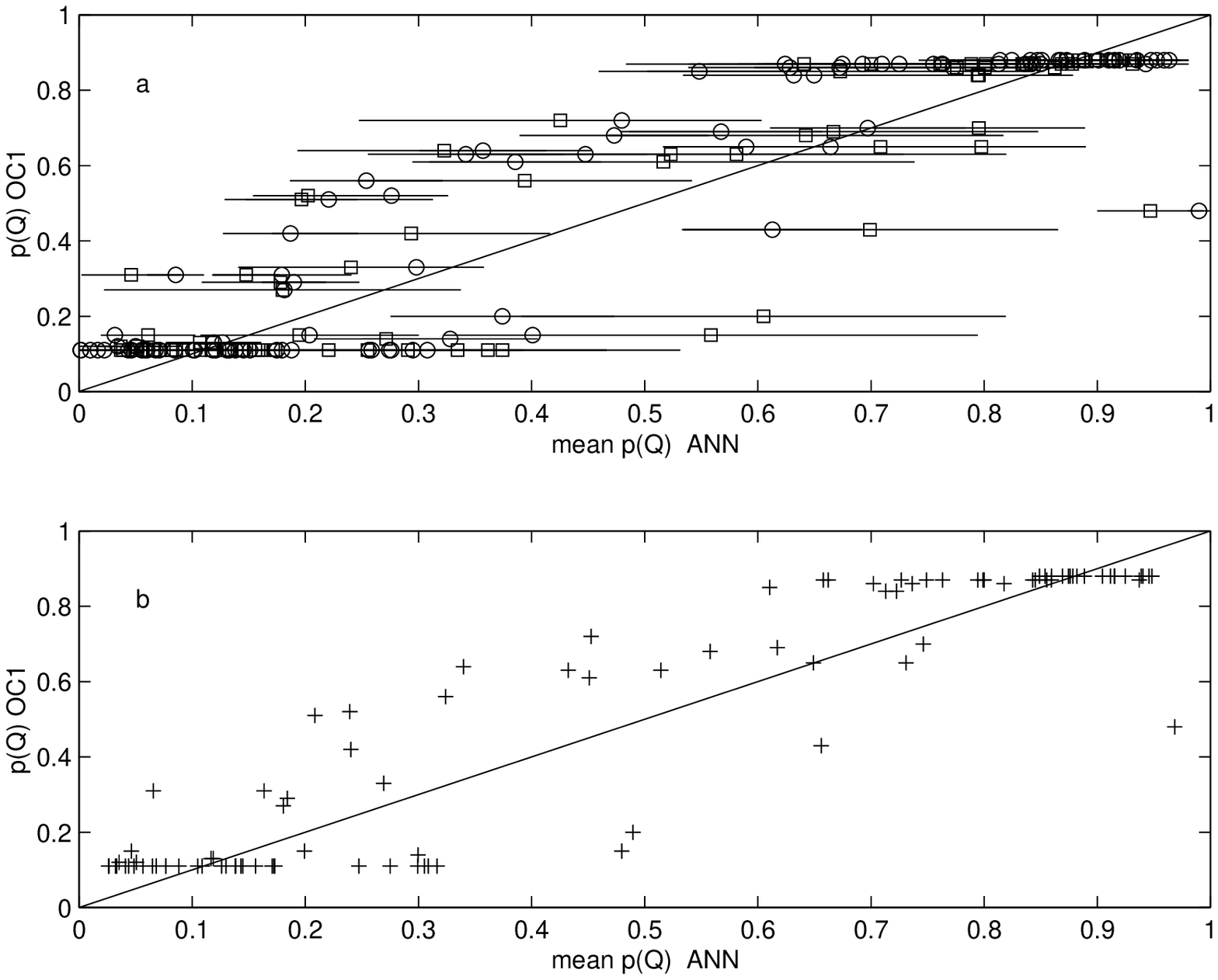,height=6.6cm } \caption{{\bf a}: $p$(Q)
OC1 versus $\overline{p}$(Q) for the 7:1 (circles) and 7:2:1
(squares) ANN models. {\bf b}: $p$(Q) OC1 versus the average of
$\overline p$(Q) for the 7:1 and the 7:2:1 ANN models.}
\end{figure}

Fig. 14  shows $\overline  p$(Q) for the  7:2:1 model versus
$\overline  p$(Q) for the 7:1 model. The agreement between the two
predictions is generally very good, with the largest discrepancies
occurring at intermediate probabilities, where the candidates'
parameters do not  fit the quasar class  nor  the nonquasar class.
The average difference between $\overline p$(Q) for the two models
(7:2:1 $-$ 7:1)  is of only 0.04,  and the standard deviation of
the difference is 0.06.

Fig. 15a shows $p$(Q) for  OC1 versus $\overline p$(Q) for the two
ANN models. Fig. 15b  is a similar  plot in which  the abcissa
corresponds  to the average of $\overline  p$(Q) for the two ANN
models. The mean and standard deviation  of the differences are
(0.04, 0.14) for OC1$-$7:1, (0.005, 0.13) for OC1$-$7:2:1 and
(0.02, 0.13) for OC1 minus the average of the two  ANN models.
Figs. 14 and 15, and the standard deviation values, show that the
agreement in $\overline p$(Q) between the two ANN models is better
than the agreement between any of them (or their average) and OC1.

The 98  sources in  Table 4 were sought for  associations in  the
NASA Extragalactic Database (NED).   Five  of them  are confirmed
extragalactic sources  with spectroscopic   redshift. FBQS
J091309.2+413635 and    FBQS J125018.1+364914 are classified as
Ultraluminous Infrared Galaxies with  redshifts 0.22 and 0.279 in
Stanford, Stern  \& de  Breuck (2000). The authors state that the
majority  of the ULIRGs  in their sample are star forming
galaxies,  and   this interpretation is consistent with the low
quasar  probability we found, of $\sim$ 0.07 and  $\sim$ 0.105
respectively (0.31 and 0.11 with OC1).  FBQS J120354.7+371137,
with $z = 0.401$, has broad emission  lines (Appenzeller et al.
1998), therefore corresponds to the  quasar classification in our
study, and it has in fact a quasar probability $\sim$ 0.94 (0.88
for OC1). A similar case is FBQS J125142.2+240435, with $z  =
0.188$ and broad emission lines (Chen  et  al. 2002),  and  also a
high quasar probability, $\sim$ 0.82 (0.86  for OC1). FBQS
J153411.3+262124 has $z=0.1294$ and spectral type ``possibly
Seyfert'' (Keel, de Grijp \& Miley 1988). We measure for the
source $p$(Q) $\sim$ 0.04 (OC1 gives 0.11), which favours a
spectral type Seyfert2, of narrow  emission lines. In addition,
four objects in Table 4 have spectroscopic classification in the
recent Sloan Digital Sky Survey Data Release 2 (SDSS DR2). FBQS
J083522.7+424258, FBQS J153402.2+425249 and FBQS J164733.9+364055
are quasars at $z = 0.805$, $z=0.649$ and $z=1.566$, with $p$(Q)
$\sim$ 0.71, $\sim$ 0.88 and $\sim$ 0.73 (0.86, 0.88 and 0.87 with
OC1). FBQS J074342.2+321543 is a star with $p$(Q) $\sim$ 0.48
(0.15 for OC1). Summarizing these results, the inspection of NED
and SDSS DR2 shows that the five FBQS candidates classified as
quasars have in fact rather high quasar probabilities - $\sim$
0.94, $\sim$ 0.82, $\sim$ 0.71, $\sim$ 0.88 and $\sim$ 0.73 -,
whereas the two ULIRGs and the star have $p$(Q) values $\sim$
0.07, $\sim$ 0.105 and $\sim$ 0.48,  reinforcing the high
efficiency of the ANN models.

\section{Conclusions}

In this work we analyse the performance of neural networks for the
selection  of quasar  candidates from  combined radio and optical
surveys with  photometric and morphological data. Our work is
based on the candidate list leading to FBQS-2 (White et al. 2000),
and the  input parameters  used are  radio flux, integrated to
peak flux ratio, photometry and point spread function in the red
and blue bands, and radio-optical position separation.

Two ANN architectures were investigated:  a logistic model (7:1)
and a model with a  hidden layer with two nodes  (7:2:1), and both
yielded similarly  good performances,  allowing  to obtain
subsamples of quasar candidates from  FBQS-2 with efficiencies as
large  as 87\% at 80\% completeness. For comparison  the quasar
fraction from the original candidate  list  was  56\%. More
complex architectures were not explored, since the inclusion of
the hidden layer - increasing the free parameters from 8 to 19 -
did not improve the performance of the network. The efficiencies
we find for completeness in  the range 70  to 90\% are 90--80\%,
similar to those found by  White et al. using  the oblique
decision tree classifier OC1 and a similar sample size for the
training. The lack of a clean separation between quasars and
nonquasars in the parameter space certainly limits the accuracy of
the classification, and the agreement in the performances obtained
favours in fact the interpretation that the three classifiers
approach the maximum value achievable with this database. Although
none of the two artificial intelligence tools provides a secure
quasar classification (say efficiency larger than 95\% for a
reasonable completeness), they are powerful to prioritize targets
for observation.

We report the  probabilities  obtained  with  the  two ANN  models
for the 98 FBQS-2 candidates  without spectroscopic classification
in White et al. Our results are compared with those found by White
et al. using OC1. The three models are found to be in agreement,
with a better match between the two ANN models (standard deviation
of the difference in probabilities $\sim$ 0.06) than between them
and OC1 (standard deviation $\sim$ 0.13).

To our knowledge, this is the first work exploring the performance
of ANNs for the selection of quasar samples. Our study
demonstrates the ability of ANNs for automated classification in
astronomical databases.

\section*{Acknowledgments}

We thank the anonymous referee for useful comments on the
manuscript. RC and JIGS acknowledge financial support from DGES
project PB98-0409-c02-02 and from the Spanish Ministerio de
Ciencia y Tecnolog\'\i a under project AYA 2002-03326. This
research has made use of the NASA/IPAC Extragalactic Database
(NED) which is operated by the Jet Propulsion Laboratory,
California Institute of Technology, under contract with the
National Aeronautics and Space Administration. Funding for the
Sloan Digital Sky Survey (SDSS) has been provided by the Alfred P.
Sloan Foundation, the Participating Institutions, the National
Aeronautics and Space Administration, the National Science
Foundation, the U.S. Department of Energy, the Japanese
Monbukagakusho, and the Max Planck Society. The SDSS is managed by
the Astrophysical Research Consortium (ARC) for the Participating
Institutions. The Participating Institutions are The University of
Chicago, Fermilab, the Institute for Advanced Study, the Japan
Participation Group, The Johns Hopkins University, Los Alamos
National Laboratory, the Max-Planck-Institute for Astronomy
(MPIA), the Max-Planck-Institute for Astrophysics (MPA), New
Mexico State University, University of Pittsburgh, Princeton
University, the United States Naval Observatory, and the
University of Washington.

\bsp

\label{lastpage}

\end{document}